\documentstyle[multicol,epsf,prl,aps]{revtex}

\def\squote{}
\def\quote#1#2#3#4{\squote {#1,\ {\sl#2}\ {\bf#3}, #4}.\par}
\def\qquote#1#2#3#4{\squote {#1,\ {\sl#2}\ {\bf#3}, #4};}

\def\prl{{\sl Phys. Rev. Lett.}\ }
\def\pr{{\sl Phys. Rev.}\ }

\def\e{\epsilon}
\begin{document}

\title{Luttinger liquid behavior in tunneling through a quantum dot
at zero magnetic field}
\author{Paula Rojt$^1$, Yigal Meir $^{1,2}$ and Assa Auerbach$^3$}
\address{$^1$Physics Department, Ben Gurion University, Beer Sheva,
84105, Israel\\
$^2$ The Ilse Katz Center for Meso- and Nanoscale Science and
Technology,  Ben Gurion University, Beer Sheva, 84105, Israel\\
$^3$ Physics Department, Technion, Haifa 32000, Israel}
\maketitle
\begin{abstract}
Thermodynamic and transport properties of a two dimensional
circular quantum dot are studied theoretically at zero magnetic
field.
 In the limit of a large confining potential,
where the dot spectrum exhibits a shell structure, it is argued
that both spectral and transport properties should exhibit
Luttinger liquid behavior. These predictions are verified by
direct numerical diagonalization. The experimental implications of
such Luttinger liquid characteristics are
 discussed.
 \end{abstract}
\begin{multicols}{2}
 Non-Fermi liquid systems - electronic
systems whose elementary excitations cannot be described by
electrons - has  fascinated physicists due to their unusual
properties (such as superconductivity and magnetism). Luttinger
Liquid (LL), describing interacting electrons in one dimension, is
one of the most studied models of such a non-Fermi liquid system,
since it has been solved long time ago \cite{luttinger}. The non
Fermi-liquid characteristics are expected to play a major role in
transport through a LL or tunneling into it \cite{kane}, giving rise to a
power-law dependence of the current or the tunneling density of
states on energy, voltage or temperature. Accordingly, there has
been many theoretical suggestions of physical systems that should
exhibit LL behavior and numerous experimental attempts to observe
this behavior in such systems. These attempts have not been very
successful, due to the sensitivity of LL to disorder, and the
difficulty of making clean one-dimensional systems. Only recently,
several decades after the  model was originally introduced, 
experimental observation of Luttinger liquid behavior
in one-dimensional systems such as nanotubes \cite{bockrath}  and
semiconductor systems \cite{amir} has been reported.
 Two dimensional systems are
also expected to exhibit LL behavior in strong magnetic fields,
due to the formation of one dimensional edge-states near the edge
of the systems \cite{wen}. However, tunneling experiments into the
edge of a two dimensional electron gas in strong magnetic fields
\cite{chang} have not been conclusive.

Quantum dots have been used in the last several years to study
various aspects of strongly correlated systems, most notably the
Kondo effect \cite{kastner,goldhaber}.  The experimental control over the
properties of the dot allow a more detailed investigation of the
strongly correlated states than that accessible in bulk systems.
An experimental study of LL behavior in quantum dots has been suggested
 by Kinaret et al.
\cite{kinaret} who studied theoretically tunneling into a quantum
dot in a strong magnetic field. Here, due to the finite number of
electrons in the dot $N$, the power-law behavior of the tunneling
density of states as a function of energy, characteristic of LL,
will be manifested in the power-law dependence of the tunneling
probability on $N$.

Technological progress allows now a construction of circularly symmetric
quantum dots \cite{tarucha}. The spectrum of these
dots displays
 shell structure expected from such symmetry.
In this letter we propose and demonstrate a new type of Luttinger
liquid behavior in such circular two-dimensional quantum dots at zero
magnetic field, which is manifested in the spectral and 
transport properties
 of the dot. The main idea in this work is that 
because of the
shell structure, one of the two quantum numbers needed to characterize an
electronic state in two dimensions is frozen (for excitations
within the shell), and thus the excitations are described by an
effective one-dimensional system. We explicitly solve the quantum
dot Hamiltonian with long-range electron-electron
interactions and obtained the
low-energy spectrum and wave-functions of this many-body system.
The spectrum and the tunneling amplitude indeed follow the
predictions of Luttinger liquid behavior.

We describe a quantum dot by a system of ${\cal N}$ interacting
spinless electrons in two dimensions, confined  by a parabolic
potential,
\begin{equation}
H=-\frac{\hbar^{2}}{2m}\sum_{i=1}^{{\cal N}}{\bf \nabla}_{i}^{2}+
\frac{1}{2}m\omega_{0}^{2}\sum_{i=1}^{{\cal N}}{\bf r}_{i}^{2}
+\sum_{i<j}V(|\bf{r}_{i}-\bf{r}_{j}|),
\label{H}
\end{equation}
with the interaction potential $V(r)$. The noninteracting part of
the Hamiltonian is trivially diagonalized, and its eigenstates and
energies are given by
\begin{eqnarray}
 \Psi_{n,\ell}(r)&=&A r^{|\ell|}e^{-r^{2}/2+\imath\theta \ell}\
\!_{1}F_{1}(-(n-|\ell|-1)/2, |\ell|+1,r^{2}) \nonumber\\
E_{n,\ell}&=& n\hbar\omega_{0},
\end{eqnarray}
with $n=1,2,\ldots$ and $\ell=-(n-1),-(n-3)\ldots,n-1$, and where
$_{1}F_{1}$ is the confluent hypergeometric function. $r$ and
$\theta$ are, respectively,  the radial and azimuthal coordinates
of the two-dimensional vector ${\bf r}$, and $r$ is in units of
$\sqrt{\hbar/m \omega_0}$. Thus the spectrum consists of equally
spaced shells. The energy of the $n$-th shell is
$n\hbar\omega_{0}$, and its degeneracy is $n$. Within a given
shell the states are characterized by a single quantum number,
e.g. the angular momentum $\ell$.

The average mean-field Coulomb interactions between electrons
leads to the usual Coulomb blockade, but does not affect the
spectral properties for a given electron number ${\cal N}$. For the
rest of the paper, while including the full Coulomb interaction
 (the last term in (\ref{H})), we make the assumption that the
inter-shell gap $\hbar\omega_{0}$ is significantly larger than the
{\sl fluctuations} in the Coulomb interactions between electrons
in different levels, which are expected to be small
\cite{altshuler}. Thus the Coulomb interactions cannot excite
electrons from one shell to another. Consider now a quantum dot
containing ${\cal N}$ electrons occupying $n_{sh}$ shells. All
shells except the last will be fully occupied, leaving
$N={\cal N}-n_{sh}(n_{sh}-1)/2$ electrons in the partially occupied
shell (POS). Since the Coulomb interaction do not excite electrons
between the shells, then the ground state and the lowest excited
states will belong to a subspace containing $n_{sh}\choose N$
states,  each consists of ${\cal N}-N$ electrons fully occupying
the lower $n_{sh}-1$ shells and $N$ electrons occupying $N$ of the
$n_{sh}$ states in the POS. All these states are degenerate in the
noninteracting limit. Our aim is to diagonalize the full
interacting Hamiltonian (\ref{H}) in this subspace.

In principle, the Coulomb interactions scatter two electrons from
two initial states onto two final states, with the conservation of
angular momentum. In the absence of inter-shell excitations, these
terms  can be divided into three contributions - (a) those where
the two initial states (and thus the
final states) are in the filled shells, (b)  those where one of
the initial states is in the filled shell (and is necessarily one
of the final states), and (c)  those where the two states are in
the POS. In the relevant subspace the first contribution is a
constant, and thus is disregarded. The second type of terms, the
effective (Hartree) potential due to the filled shells \cite{exchange}, 
contribute to the diagonal of the Hamiltonian matrix,
 leading to the removal of the
energy degeneracy between the different angular momenta states
(angular momenta $\pm \ell$ are still degenerate). The resulting
dispersion turns out to be  very close to an inverse parabola.

Thus, the effective Hamiltonian in the relevant subspace can be
written as
\begin{equation}
{\cal H}_{eff} = \sum_\ell \e_\ell c^\dagger_\ell c_\ell +
\sum_{j,k,\ell} V_{k,j,\ell} c^\dagger_{\ell-k} c^\dagger_{j+k} c_j
c_\ell \label{Heff}
\end{equation} with
\begin{eqnarray}
 \e_\ell=\sum_{n<n_{sh},j} \int\int d^2r'd^2r|\Psi_{n,j}(r)|^2V(r-r'&)&|\Psi_{n_{sh},\ell}(r')|^2,
 \nonumber\\
V_{k,j,\ell}=\int\int d^2r d^2r'
\Psi_{n_{sh},\ell-k}^*(r')\Psi_{n_{sh},j+k}^*(r&)&\nonumber\\
\times V(r-r')\Psi_{n_{sh},j}(r)\Psi_{n_{sh},\ell}&(&r')
\label{Hqd}
\end{eqnarray}
where $c_\ell$ annihilates a particle with angular momentum $\ell$
in the POS, and  the summations in ${\cal H}_{eff}$ are over all
angular momenta states in the POS.

This effective Hamiltonian in the POS resembles the Hamiltonian
describing  interacting electrons in one dimension,
\begin{equation}
{\cal H}_{1d} = \sum_k \e_k c^\dagger_k c_k + \sum_k V_{k} \rho_{k}
\rho_{-k}
\label{H1d},
\end{equation}
where $\rho_k\equiv\sum_j c^\dagger_{k+j} c_k$, $\e_k=v_F k$, with
$k$ the momentum quantum number and $v_F$ the Fermi velocity.
Comparing the effective Hamiltonian (\ref{Heff}) to (\ref{H1d}) we
note that in our case the interaction $V$ depends on three
quantum numbers, and not only on their difference. Thus unlike
the LL model, the interaction is not merely a product of densities.
 This difference arises because the system is
not strictly one dimensional, but the electronic wave-functions in
the $n_{sh}-th$ shell, while having the same value of $\langle
r^2\rangle\propto n_{sh}$, do have a nontrivial dependence on $r$.
Additionally, the number of the electrons in the POS, which are
described the the effective Hamiltonian (\ref{Heff}), is finite
(and could be small).

Since the one-dimensional Hamiltonian (\ref{H1d}) is exactly
solvable and is described by LL theory, it is tempting to check
whether the two-dimensional behavior of a symmetric quantum dot is
governed by LL physics. In order to investigate the relevance of
LL behavior to the spectral and transport properties of the dot,
we diagonalized numerically the Hamiltonian (\ref{Heff}) for
several values of $n_{sh}$ and $N$, and for a particular choice of
interaction, $V(r) = \log(1/|r|)$. This choice allows us to
evaluate all the matrix elements $\e_\ell$ and $V_{j,k,\ell}$
analytically. To investigate the particular role of the
interaction term (the second term in (\ref{Heff})), we multiplied
it by a prefactor $\alpha$ and carry out our study for several
values of $\alpha$. The results shown here are for $n_{sh}=20$.
For each value of total angular momentum $L$, the Hamiltonian was
diagonalized in the subspace of the lowest 6
eigenstates\cite{error}, in order to determine the lowest energy
state with angular momentum $L$.

For an even number of electrons,
the lowest energy state is of total angular momentum $L=0$, the first
excited states carry angular momentum $L=\pm2$, then $L=0,\pm4$, etc.
According to LL theory (for spinless electrons),
 the low-energy spectrum of (\ref{H1d}) is
given by \cite{mahan}
\begin{equation}
E_p = p \sqrt{v_F^2+2 v_F V_p/\pi} \label{gap}
\end{equation}
\noindent
In order to check the spectrum of the two-dimensional quantum dot
(\ref{Hqd}) against the predictions of LL theory,
 we compare in Fig.~1 the
evaluated gap, $\Delta\equiv(E_{L=2}-E_{L=0})/2$, to the
expectation from LL theory (\ref{gap}). As can be clearly seen
from the figure, the nontrivial dependence of the gap on the interaction
 strength, $\alpha$, is very well fitted by the functional
form expected from LL theory, with a single fitting parameter
$V_{p=2}=0.055$. (Note that in our case the Fermi velocity, $v_F$,
is also $\alpha$-dependent, due to the diagonal interaction
terms.) Numerically we indeed find that the interaction integrals
$V_{2,j,\ell}$ are about $0.05\pm0.01$ for several values of $k$
and $\ell$ (independent on what branch $k$ and $l$ are). The
interaction integrals for higher values of $p$ are much smaller,
e.g. $V_{4,j,\ell}\simeq 0.01$. For $\alpha=20$, however, we see
deviations from the theoretical predictions, indicating perhaps
that such strong interactions involve states beyond the linear
dispersion  regime.


\begin{center}
\leavevmode \epsfxsize=3.3in
\epsfbox{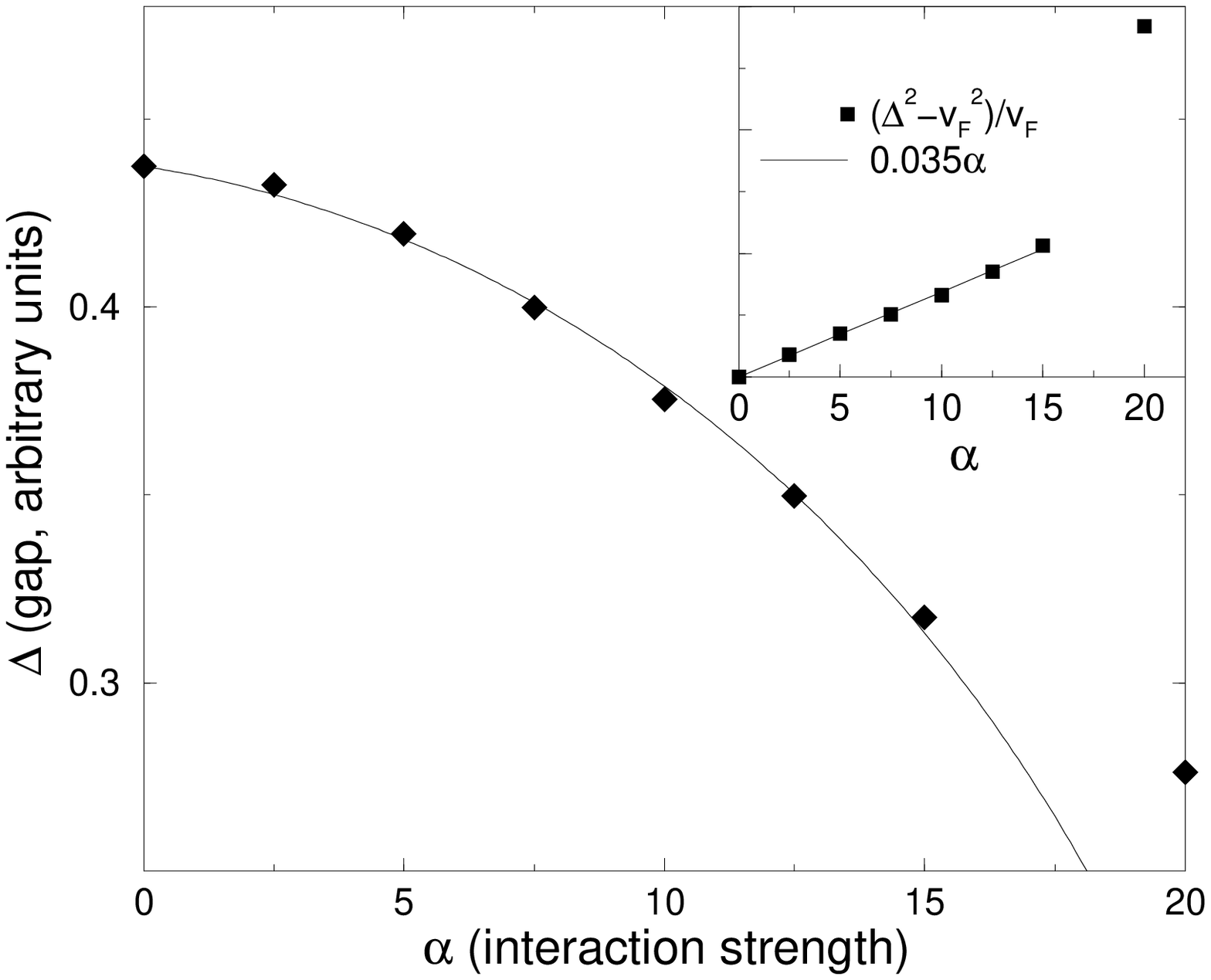}
\end{center}
\begin{small}
\vskip -0.3 truecm Fig. 1. Comparison of the numerically evaluated
gap $\Delta$ in the quantum dot spectrum to the prediction of Luttinger
liquid theory [Eq.(\ref{gap})]. There is a single fitting
parameter, $V_{p=2}=0.055$ (see text). Inset: Plot of $(\Delta^2-v_F^2)/v_F$
vs. the interaction strength. Luttinger liquid theory predicts a straight
line, of slope $2V_2/\pi$.
\end{small}
\vskip 0.5 truecm

LL characteristics are manifested not only in the spectrum, but
also in the matrix elements, or the tunneling probabilities. The
amplitudes of the Coulomb blockade peaks, $T_N$, in the
 linear response conductance through the quantum dot are proportional,
 when the temperature
 is smaller than the gap in the excitation spectrum,  to \cite{ourlandauer}
\begin{equation}
T_N \propto |<\Psi_N| d_\ell|\Psi_{N+1}>|^2 \ ,
\end{equation}
where $\Psi_N$ is the many-particle ground state of a system with
$N$ particles is the POS, and $d_\ell$ annihilated a particle with
the corresponding angular momentum (namely, the difference in
total angular momentum between $\Psi_N$ and $\Psi_{N+1}$). This
quantity was studied extensively for the case of quantum dot in
the fractional quantum Hall regime by Kinaret et.
al.\cite{kinaret}. Using the explicit form of the edge-state
propagators they showed that $T_N \propto N^{-(1/\nu-1)/2}$,
 where $\nu$ is the filling factor. This dependence was also
 verified numerically. The fractional quantum Hall liquid
  turned out to be a special case
 of the general theory of tunneling into a LL\cite{kane}, where it
 was shown that the tunneling probability depends on energy in
 a power-law manner,
 \begin{equation}
 T_N \propto \e^{\beta-1},\ \ \ \beta=(g+g^{-1})/2
 \label{powerlaw}
 \end{equation}
where $g$ characterizes the strength of interactions in LL theory
($g=1$ no interactions, $g<1$ repulsive interactions). The
power-law dependence on $\e$ is directly translated into a
power-law dependence of $T_N$ on $N$, $T_N \propto N^{-(\beta-1)/2}$
(the factor $2$ comes from the two-dimensionality,
$N\sim L^2\sim k^{-2}\sim\e^{-2}$).

To check these LL predictions we have evaluated numerically the
matrix elements $|<\Psi_N| d_\ell|\Psi_{N+1}>|^2$, for the same
set of parameters mentioned above, and for different values of the
interaction strength $\alpha$. As can be seen in Fig.~2, these
matrix elements exhibit power-law dependence on the number of
particles in the POS, in agreement with LL theory. Surprisingly
the power-law behavior is already manifested in tunneling into a
quantum dot with a small number of electrons in the POS.

\begin{center}
\leavevmode \epsfxsize=4.1in
\epsfbox[32 -38 596 426]{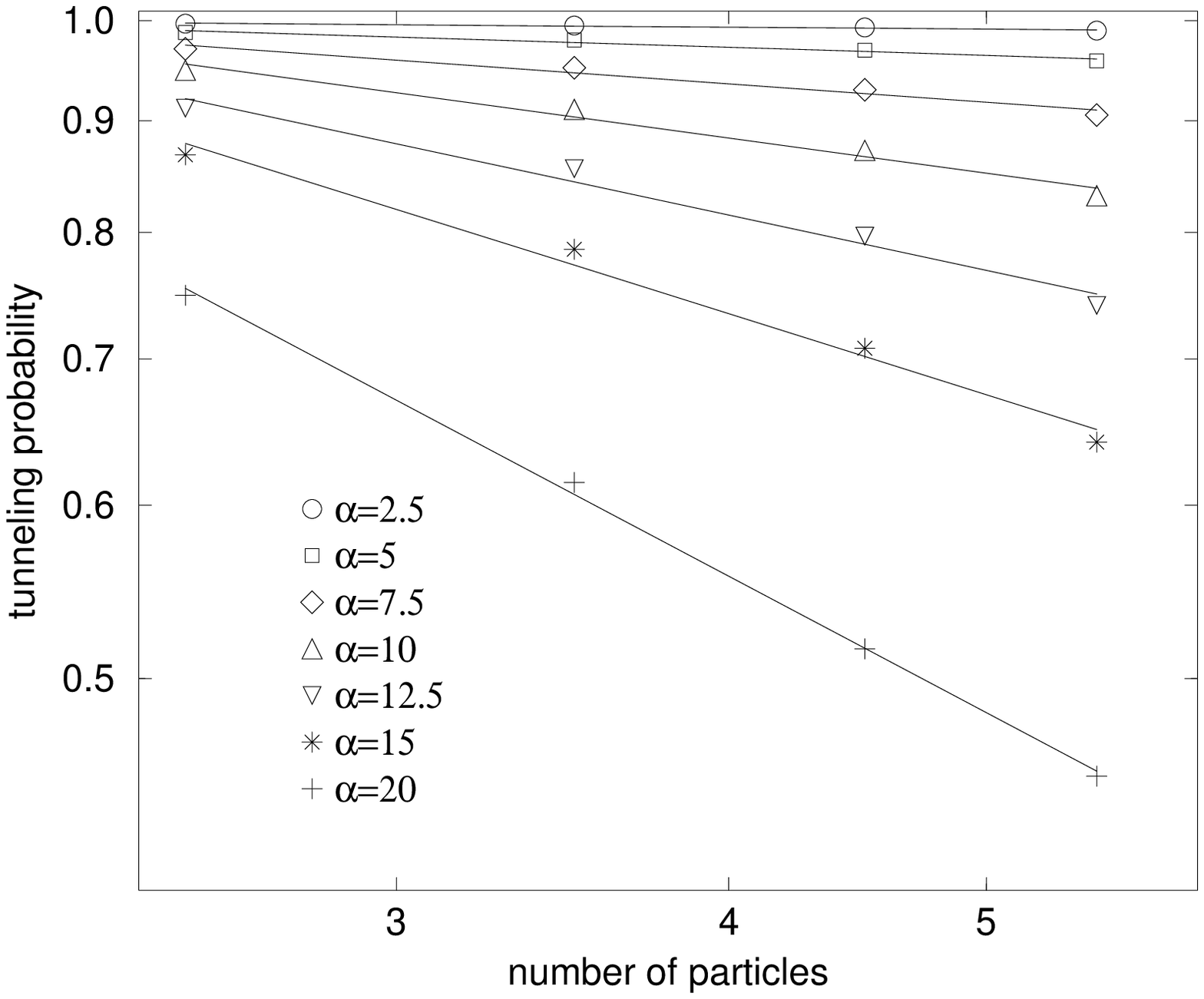}
\end{center}
\begin{small}
\vskip -1.8 truecm Fig. 2. The numerically evaluated tunneling
matrix elements $T_N$, which are proportional to the Coulomb
blockade peak amplitudes, for several values of the interaction
strength $\alpha$. In agreement with Luttinger liquid theory, a
power-law dependence of $T_N$ on $N$ is observed, with an exponent
that increases with the strength of the interactions.
\end{small}
\vskip 0.5 truecm

To further check the relevance of LL theory to the tunneling
matrix elements, we employ the functional dependence of the LL
parameter $g$ on the interaction strength, which can be explicitly
evaluated for a one-dimensional model with long-range interactions
(only $V_0\ne0$ in (\ref{H1d})). For this case \cite{mahan},
\begin{equation}
g = {1\over{\sqrt{1+V_0/\pi v_F}}} \label{g}
\end{equation}
In Fig.~3 we compare the resulting exponent to the prediction of
LL theory with such long range interactions. We see an excellent
agreement between the numerically obtained exponent and the
theoretical prediction, with a single fitting parameter $V_0\simeq0.33$ 
(with again a deviation at very large interaction strength, $\alpha=20$).
 The agreement
 between the theoretical prediction and the numerical calculation
also indicates that the logarithmic interactions used in the
numerics mimic the long-range nature ($V_p=0, p>0$) of the
prediction (\ref{g}).

\begin{center}
\leavevmode \epsfxsize=3.3in
\epsfbox{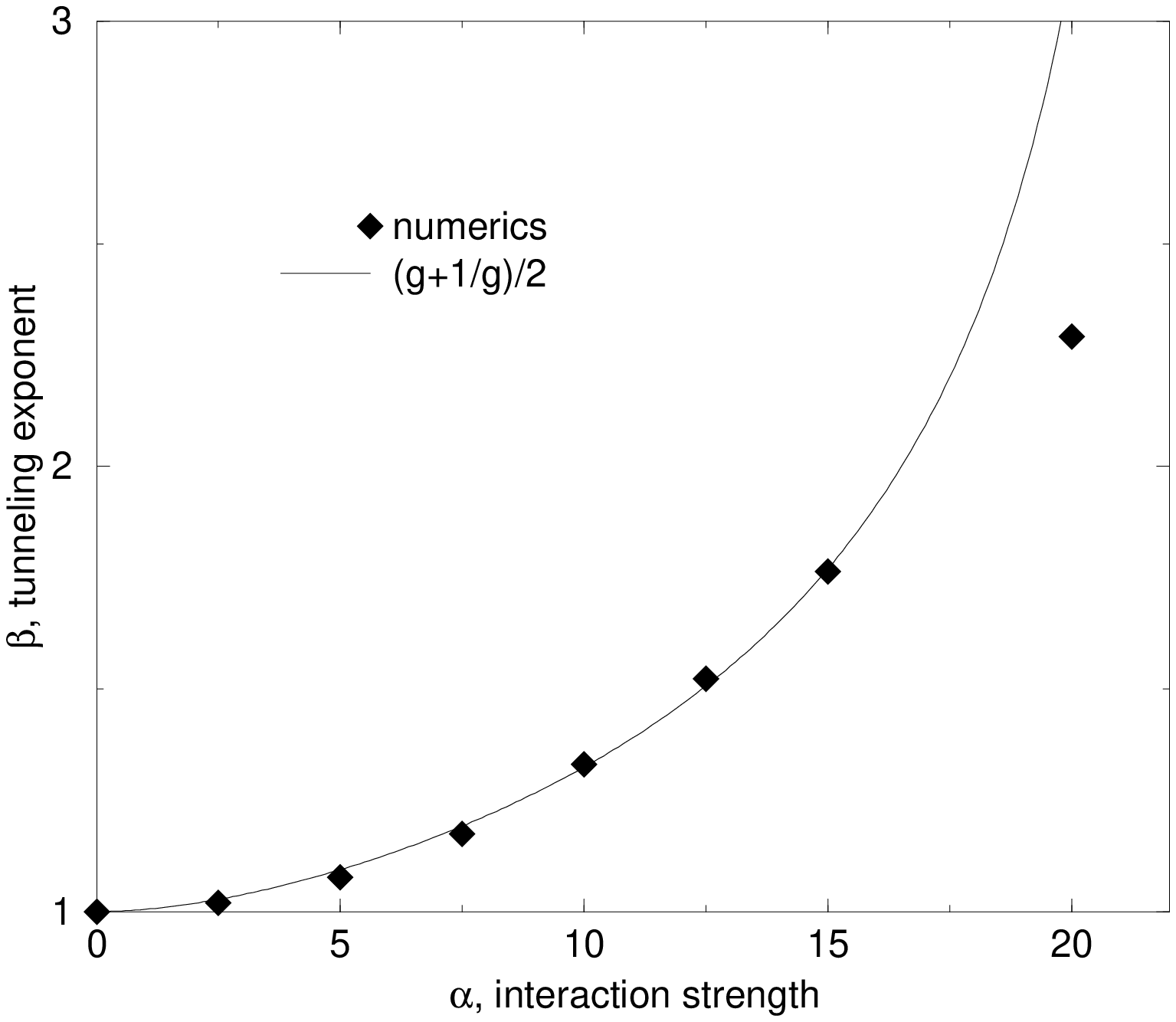}
\end{center}
\begin{small}
\vskip -0.3 truecm Fig. 3. The numerically evaluated tunneling
matrix exponent $\beta$, derived from Fig.2, as a functions of
the interaction strength, $\alpha$. The dependence upon $\alpha$
exhibits a good agreement with the predictions of Luttinger liquid
theory, Eqs.(\ref{powerlaw},\ref{g}).
\end{small}
\vskip 0.5 truecm

To conclude, we have demonstrated, by heuristic arguments and by
numerical calculations, that the excitations in a circular quantum
dot at zero magnetic field are described by Luttinger liquid
theory. These should have an observable influence on transport
through quantum dots. We predict that in a given shell the
amplitude of the Coulomb blockade peak decreases in a power-law
fashion. Taking into account that this power-law was observed
numerically for a quantum dot with a small number of electrons in
the outer shell (see Fig.2),  this prediction can be tested with
existing structures.

An even more intriguing possibility is the relevance of this work
to  transport through a system of interacting electrons in the
presence of potential disorder,  one of the most
challenging problems of condensed matter physics.   When the
disorder potential is large enough to break the electron liquid
system into puddles, or effective quantum dots, the overall
dissipative resistance may be determined by transport through
such a random quantum dot array \cite{xi}. Thus the present 
calculation suggests that transport in a two-dimensional 
disordered interacting system may be dominated by LL behavior, 
similar to transport in such a system in a strong magnetic
field \cite{shimshoni}.

In this paper the spin of the electrons was ignored. It would be
quite interesting to investigate the role of spin in such devices and
explore the possibility of e.g. spin-charge separation and its relevance
to transport through a quantum dot.
Another avenue worth exploring is the question whether other systems exhibiting
shell structure, such as nuclei and atoms, can also, under some circumstances,
exhibit Luttinger liquid behavior. These questions will be investigated in a
future publication.

Work at BGU was supported by DIP.
AA acknowledges support from  US-Israel Binational
Science Foundation and the Fund
For Promotion Of Research at Technion.

\end{multicols}


\begin{references}
\bibitem{luttinger}
\quote{S. Tomonaga}{Prog. Theor. Phys. (Kyoto)}{5}{544 (1950)}
\qquote{J. M. Luttinger}{J. Math. Phys. N.Y.}{4}{1154 (1963)}

\bibitem{kane}
\qquote{C. L. Kane and M. P. A. Fisher}{\prl}{68}{1220 (1992)}
 {\sl \pr} {\bf B 46}, 15233 (1992).

\bibitem{bockrath}
\qquote{M. Bockrath et al.}{Nature}{397}{598 (1999)} \qquote{Yao
et al.}{Nature}{402}{273 (1999)}

\bibitem{amir}
\quote{O. Auslaender et al.}{\prl}{84}{1764 (2000)}

\bibitem{wen}
\quote{X.-G. Wen}{\pr}{B 41}{12838 (1991)}

\bibitem{chang}
\qquote{A. M. Chang et al.}{\prl}{77}{2538 (1996)}
\qquote{M.
Grayson et al.}{\prl}{80}{1062 (1998)} \qquote{A. M. Chang et
al.}{\prl}{86}{143 (2001)}
 \quote{M. Grayson et al.}{\prl}{86}{2645 (2001)}

\bibitem{kastner}
For a review on transport through quantum dots see, e.g.,
\quote{M. A. Kastner}{Comments Condens. Matter Phys.}{17}{349 (1996)}

\bibitem{goldhaber}
\qquote{D. Goldhaber-Gordon et al.}{Nature}{391}{156 (1998)}
\qquote{S. M. Cronenwett et al.}{Science}{281}{540 (1998)}
\qquote{J. Schmid et al.}{Physica}{B 256-258}{182 (1998)}
\quote{P. Simmel et al.}{\prl}{83}{804 (1999)}


\bibitem{kinaret}
\qquote{J. M. Kinaret et al.}{Phys. Rev.}{B 45}{9489 (1992)}
\quote{J. M. Kinaret et al.}{Phys. Rev.}{B 46}{4681 (1992)}

\bibitem{tarucha}
\qquote{S.
Tarucha et al.}{\prl}{77}{3613 (1996)}
\qquote{L.~P. Kouwenhoven et al.}{Science}{278}{1788 (1997)}

\bibitem{altshuler}
\quote{L. L. Kurland, I. L. Aleiner and B. L. Atshuler}
{\pr}{B 62}{14886 (2000)}

\bibitem{exchange}
Since we are dealing with spinless electrons, exchange terms have been
ignored in this paper.

\bibitem{error}
By comparing to the full diagonalization for a small number of
electrons, we can estimate the error in the minimal energy  for a
given angular momentum due to  the subspace reduction to be less
than 1\%.

\bibitem{mahan}
See, e.g., G. D. Mahan, {\sl Many-Particle Physics} (Plenum, New York, 1990).

\bibitem{ourlandauer}
\quote{Y. Meir and N. S. Wingreen}{\prl}{68}{2512 (1992)}

\bibitem{xi}
\qquote{Song He and X.C. Xie}{Phys. Rev. Lett.}{80}{3324 (1998)}
\qquote{Y. Meir}{Phys. Rev. Lett.}{83}{3506 (1999)}
\qquote{J. Shi, Song He, and X.C. Xie}{Phys. Rev.}{B 60}{R13950 (1999)}
\qquote{Y. Meir}{Phys. Rev.}{B. 61}{16470 (2000)}
{\sl ibid} {\bf 63}, 073108 (2001).

\bibitem{shimshoni}
\qquote{E. Shimshoni and A. Auerbach}{Phys. Rev.}{B 55}{9817 (1997)}
\qquote{E. Shimshoni, A. Auerbach, and A. Kapitulnik}{Phys. Rev. Lett.}{80}
{3352 (1998)}
\quote{E. Shimsoni}{Phys. Rev.}{B 60}{10691 (1999)}

\end{references}
\end{document}